\documentstyle[aps,prd]{revtex}
\def\be{\begin{equation}}
\def\ee{\end{equation}}
\def\bea{\begin{eqnarray}}
\def\eea{\end{eqnarray}}
\def\s{\sigma}
\def\lm{\lambda}
\def\de{\delta}
\def\om{\omega}
\def\pr{\prime}
\begin{document}
\wideabs{ 
\title{Quasirotational Motions \\
and Stability Problem in Dynamics\\
of String Hadron Models}
\author{G. S. Sharov}
\address{
Tver State University,
Sadovyj per., 35, 170002, Tver, Russia}
\maketitle

\begin{abstract}

For the relativistic string with massive ends (the meson model)
and four various string baryon models ($q$-$qq$, $q$-$q$-$q$, Y and
$\Delta$) we consider the classical quasirotational motions, which
are small disturbances of the planar uniform rotations
of these systems. For the string meson model the two types of these solutions
are obtained. They describe oscillatory motions in the form of stationary
waves in the rotational plane and in the orthogonal direction.
This approach and the suggested method of determining an arbitrary motion
of the system on the base of initial data let us solve the stability problem
for the rotational motions for all mentioned string configurations.

It is shown that the classic rotational motions are {\it stable} for the
string meson model (or its analog $q$-$qq$) and for the $\Delta$ baryon
configuration, but they are {\it unstable} for the string baryon models
Y and $q$-$q$-$q$.
For the latter two systems any small asymmetric disturbances grow with growing
time. The motion of the  $q$-$q$-$q$ configuration become more complicated and
quasiperiodic but quarks do not merge. In the case of Y model the evolution of
disturbances results in falling a quark into the junction.
These features of the classic behavior are important for describing the hadron
states on the Regge trajectories and for choosing and developing
the most adequate QCD-based string hadron models.

\end{abstract}
} 

\section*{Introduction}

The string hadron models \cite{Nambu,Ch,BN,Ko,lin,AY,PY,Tr,PRTr}
use the striking analogy
between the QCD confinement mechanism at large interquark distances and
the relativistic string with linearly growing energy
connecting two material points. Such a string with massive ends may be
regarded as the meson string model \cite{Ch,BN}.

String models of the baryon were suggested in the following four variants
(Fig.~1) differing from each other in the topology of spatial junction
of three massive points (quarks) by relativistic strings:
(a) the quark-diquark model $q$-$qq$\cite{Ko} (on the classic level
it coincides with the meson model \cite{Ch});
(b) the linear configuration $q$-$q$-$q$\cite{lin};
(c) the ``three-string" model or Y-configuration \cite{AY,PY}
and (d) the ``triangle" model or $\Delta$-configuration \cite{Tr,PRTr}.

\begin{figure}[bh]
\unitlength=0.6mm
\begin{center}
\begin{picture}(154,28)
\thicklines
\put(7,15){\line(1,0){25}}
\put(7,15){\circle*{2}}\put(32,16){\circle*{2}}\put(32,14){\circle*{2}}
\put(6,10){$q$}\put(29,9){$qq$}\put(11,26){a}
\put(47,15){\line(1,0){28}}
\put(47,15){\circle*{2}}\put(61,15){\circle*{2}}\put(75,15){\circle*{2}}
\put(46,10){$q$}\put(60,10){$q$}\put(74,10){$q$}\put(52,26){b}
\put(97,15){\line(0,-1){12}}\put(97,15){\line(-2,1){10}}
\put(97,15){\line(2,1){10}}
\put(97,3){\circle*{2}}\put(87,20){\circle*{2}}\put(107,20){\circle*{2}}
\put(93,3){$q$}\put(83,17){$q$}\put(108,17){$q$}\put(93,26){c}
\put(115,5){\line(1,0){24}}\put(115,5){\line(2,3){12}}
\put(127,23){\line(2,-3){12}}
\put(115,5){\circle*{2}}\put(139,5){\circle*{2}}\put(127,23){\circle*{2}}
\put(111,4){$q$}\put(141,4){$q$}\put(123,24){$q$}\put(132,26){d}
\end{picture}\end{center}
\caption{String baryon models.}
\end{figure}
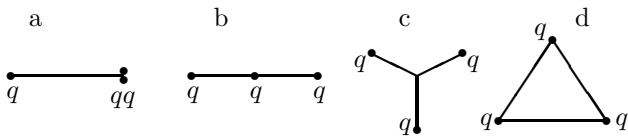

The problem of choosing the most adequate string baryon model among
the four mentioned ones has not been solved yet.
Investigation of this problem from the point of view of the QCD limit
at large distances has not been completed. In particular,
the QCD-motivated baryon Wilson loop operator approach
gives some arguments in favour of the Y-configuration \cite{Isg} or
the ``triangle" model \cite{Corn}. Leaving the Wilson loop analysis outside
this paper we concentrate on the classical dynamics of these configurations.

For all mentioned string hadron models the classical solutions
describing the rotational motion (planar uniform rotations of the
system) are known and widely used for modeling the orbitally excited hadron
states on the main Regge trajectories \cite{Ko,Solov,4B,InSh}.
The rotational motion of the meson model or the baryon configuration
$q$-$qq$ is a rotation of the rectilinear string segment \cite{Ch,Ko}.
For the model $q$-$q$-$q$ the motion is the same but with the middle quark
at the rotational center.
The form of rotating three-string configuration is three rectilinear
string segments joined in a plane at the angles 120${}^\circ$ \cite{AY,PY}.
For the model ``triangle" this form is the rotating closed
curve consisting of segments of a hypocycloid \cite{Tr,PRTr}.

In this paper we consider the motions of all these systems which are close to
rotational ones. They are interesting due to the following two reasons:
(a) we are to search the motions describing the hadron states, which are
usually interpreted as higher radially excited states in the potential models \cite{InSh}, in other words, we are to describe the daughter Regge trajectories;
(b) the important problem of stability of rotational motions has not been
solved yet for all mentioned string models. For the meson string model
the quasirotational motions of slightly curved string with massive ends
were searched in Ref.~\cite{AllenOV}. But some simplifying
assumptions in these papers oblige us in Sect.~\ref{sec1} to verify the
solutions \cite{AllenOV} numerically and to suggest another approach for
obtaining the quasirotational solutions in the form of stationary waves and
solving the stability problem. In the following Sects.~\ref{sectr}, \ref{li} and
\ref{Y} the stability problem for the string baryon configurations $\Delta$,
$q$-$q$-$q$ and Y (Fig.~1) is tested with using the suggested method of solving the initial-boundary value problem for these systems.

\section{String with massive ends}
\label{sec1}

The relativistic string with the tension $\gamma$ and the masses
$m_1$, $m_2$ at the ends (the meson or $q$-$qq$ baryon model)
is described by the action \cite{Ch,BN,4B}
\be
S=-\gamma\int\limits_{\;\Omega}\!\!\!\!\int\sqrt{-g}\,d\tau\,d\s-
\sum_{i=1}^N m_i\int\sqrt{\dot x_i^2(\tau)}\,d\tau,
\label{S}\ee
written here in the general form with $N$ material points for all mentioned
hadron models. For the string with massive ends $N=2$,
$X^\mu(\tau,\s)$ are coordinates of a string point
in $D$-dimensional Minkowski space $R^{1,D-1}$ with
signature $+,-,-,\dots$, $g=\dot X^2X'{}^2-(\dot X,X')^2$,
$(a,b)=a^\mu b_\mu$ is the (pseudo)scalar product,
$\dot X^\mu=\partial_\tau X^\mu$, $X^{\pr\mu}=\partial_\s X^\mu$,
$\Omega=\{(\tau,\s):\,\tau_1<\tau<\tau_2,\;\s_1(\tau)<\s<\s_2(\tau)\}$,
$\s_i(\tau)$ are inner coordinates of the quark\footnote{We use the term
``quark" for brevity, here and below quarks, antiquarks and diquarks
are material points on the classic level.}
world lines, their coordinates in $R^{1,D-1}$ are
$x_i^\mu(\tau)= X^\mu(\tau,\s_i(\tau))$,
$\dot x_i^\mu=\frac d{d\tau} x_i^\mu(\tau)$; the speed of light $c=1$.
The first summand in Eq.~(\ref{S}) is proportional to the world surface
area and may be rewritten in the equivalent form \cite{BN,AllenOV}.

The equations of motion
\be
\frac\partial{\partial\tau}\frac{\partial\sqrt{-g}}{\partial\dot X^\mu}
+\frac\partial{\partial\s}\frac{\partial\sqrt{-g}}{\partial X^{\pr\mu}}=0,
\label{eqn}\ee
and the boundary conditions for the quark trajectories
\be
\frac d{d\tau}\frac{\dot x_{i\mu}}{\sqrt{\dot x_i^2}}-\frac{(-1)^i\gamma}{m_i}
\bigg[\frac{\partial\sqrt{-g}}{\partial X^{\pr\mu}}-\s'_i(\tau)\,
\frac{\partial\sqrt{-g}}{\partial\dot X^\mu}\bigg]\bigg|_{\s=\s_i}\!\!\!\!=0
\label{qn}\ee
are deduced from action (\ref{S}) \cite{BN}.

The exact solution of Eq.~(\ref{eqn}) satisfying conditions
(\ref{qn}) and describing the rotational motion of the rectilinear string
is well known \cite{Ch,BN,Ko} and may be represented as
\be
X^0\equiv t=b\tau,\quad
X^1+iX^2=\om^{-1}\sin(\om b\s)\cdot e^{i\om t}.
\label{rot}\ee
Here $\om$ is the angular velocity,
$\s\in[\s_1,\s_2]$, $\s_i={}$const, $\s_1\le0<\s_2$; the substitution
$\tilde\s=\sin(\om b\s)$ can be made.

The authors of Refs.~\cite{AllenOV} search motions of this system
close to the rotational one (\ref{rot}) in the form
\be
X^0=t,\quad
X^1+iX^2=\s R(t)\cdot e^{i[\om t+\phi(t)+f(\s)]},
\label{AOV}\ee
where $\phi(t)$, $f(\s)$, $\dot R(t)$ are assumed to be small.
After substituting this formula into Eq.~(\ref{eqn}) and dropping the
second order terms they obtain the expression for $f(\s)$
\be
f(\s)=\ddot\phi(t)\,\om^{-3}R^{-1}(t)\,\big[f_1(\s,v_\perp)-
f_1(1, v_\perp)\big],
\label{fAOV}\ee
where $f_1(\s,v_\perp)=(\frac12 v_\perp A+
\s^{-1}\sqrt{1-\s^2v_\perp^2})\,A$,
$A=\arcsin\s v_\perp$, $v_\perp=\om R(t)$,
$\s_1=0\le\s\le1=\s_2$, the first heavy quark with $m_1\to\infty$
is at rest.

Deducing Eq.~(\ref{fAOV}) the authors of Refs.~\cite{AllenOV} ignore that
(a) the obtained function $f(\s)$ (\ref{fAOV}) depends on $t$ essentially
[one may assume the dependencies $R(t)$, $v_\perp(t)$ weak but this can not
be valid for the multiplier $\ddot\phi(t)$] so all previous calculations
appear to be wrong; (b) expression (\ref{AOV}), (\ref{fAOV}) does not
satisfy the boundary condition (\ref{qn}) for the moving quark.
The dependencies $\phi(t)$, $f$, $R(t)-\langle R\rangle$ on time
should also be analyzed. In particular, the following important
question remains without answer: do these disturbances grow with growing time,
in other words, is the rotational motion (\ref{rot}) stable?

In our opinion, the better way of searching quasirotational motions and
solving the stability problem includes the choice of the coordinates $\tau,\s$
on the world surface (that can always be made \cite{BN}),
in which the orthonormality conditions
\be
\dot X^2+X'{}^2=0,\qquad(\dot X,X')=0,\label{ort}\ee
are satisfied.
Under restrictions (\ref{ort}) the equations of motion (\ref{eqn})
become linear
\be
\ddot X^\mu+X''{}^\mu=0,\label{eq}\ee
and the boundary conditions (\ref{qn}) take the simplest form
\be
m_i\frac d{d\tau}U^\mu_i(\tau)+(-1)^i\gamma
X^{\pr\mu}(\tau,\s_i)=0,\quad i=1,2,
\label{qq}\ee
where $U^\mu_i(\tau)=\dot x_i^\mu(\tau)\big/\sqrt{\dot x_i^2}$
is the unit $R^{1,D-1}$-velocity vector of $i$-th quark.

In Eqs.~(\ref{qq}) the functions $\s_i(\tau)$ are chosen in the form
$\s_i(\tau)={}$const. One can always fix them, in particular, as
\be
\s_1=0,\qquad\s_2=\pi
\label{fiq}\ee
with the help of the reparametrizations \cite{BN}
\be
\tilde\tau\pm\tilde\s=f_\pm(\tau\pm\s)
\label{rep}\ee
($f_\pm$ are arbitrary smooth monotone functions), which keep
invariance of Eqs.~(\ref{ort}) and (\ref{eq}).

Using the general solution of Eq.~(\ref{eq})
\be
X^\mu(\tau,\s)=\frac1{2}\big[\Psi^\mu_+(\tau+\s)+\Psi^\mu_-(\tau-\s)\big],
\label{sol}\ee
we can reduce the problem in a natural way to solving ordinary differential
equations. In particular, this approach is fruitful for considering
the initial-boundary value problem (IBVP) for the string with massive ends
\cite{BaSh}. This problem implies obtaining the motion of the
string on the base of two given initial conditions:
an initial position of the system in Minkowski space and initial
velocities of string points. In other words, we are to determine
the solution of Eq.~(\ref{eq}) $X^\mu(\tau,\s)$ satisfying the orthonormality
(\ref{ort}), boundary (\ref{qq}) and initial conditions.

An initial position of the string can be given as the parametric curve
in Minkowski space
\be
x^\mu=\rho^\mu(\lm),\qquad\lm\in[\lm_1,\lm_2],\;\quad
{\rho'}^2<0.
\label{curve}\ee
Initial velocity of a string point is a time-like vector on this curve
$\,v^\mu(\lm)$, $\lm\in[\lm_1,\lm_2]$, $v^\mu(\lm)$ may be
multiplied by an arbitrary scalar function $\chi(\lm)>0$.

To solve the problem we set parametrically the initial curve
on the world sheet
\be
\tau=\check\tau(\lm),\qquad\s=\check\s(\lm),\qquad\lm\in[\lm_1,\lm_2],
\label{tasi}\ee
and use the following general form for the initial position of the string
\cite{BaSh}:
\be
X^\mu\bigl(\check\tau(\lm),\check\s(\lm)\bigr)=\rho^\mu(\lm),
\qquad\lm\in[\lm_1,\lm_2].
\label{X=r}\ee
Here $|\check\tau'|<\check\s'$, $\check\tau(\lm_1)=\check\s(\lm_1)=0$,
$\check\s(\lm_2)=\pi$.
There is the freedom in choosing the functions $\check\tau(\lm)$,
$\check\s(\lm)$
\cite{BaSh,An}
connected with the invariance of Eqs.~(\ref{ort}), (\ref{eq}), (\ref{qq}) and
(\ref{fiq}) with respect to the substitutions (\ref{rep}) where
\be
f_+(\xi)=f_-(\xi)=f(\xi),\quad f(\xi+2\pi)=f(\xi)+2\pi
\label{MP}\ee
and $f'(\xi)>0$.

Using the formulas \cite{BaSh}
\be
\frac d{d\lm}\Psi^\mu_\pm\bigl(\check\tau(\lm)\pm\check\s(\lm)\bigr)=
\bigg[1\pm\frac{(v,\rho')}\Delta\bigg]\rho^{\pr\mu}\mp
\frac{\rho^{\pr2}}\Delta v^\mu,
\label{psi}\ee
where $\,\Delta(\lm)=\sqrt{(v,\rho')^2-v^2\rho^{\pr2}}$,
we can determine from the initial data the function $\Psi^\mu_+$
in the initial segment $\bigl[0,\check\tau(\lm_2)+\pi\bigr]$ and the function
$\Psi^\mu_-$ in the segment $\bigl[\check\tau(\lm_2)-\pi,0\bigr]$.
The constants of integration are fixed from the initial condition
(\ref{X=r}).

The functions $\Psi^\mu_\pm$ are to be continued beyond the initial segments
with the help of boundary conditions (\ref{qq}) which may be reduced
to the equations \cite{BaSh,An}
\bea
U^{\pr\mu}_1(\tau)&=&\gamma m_1^{-1}\big[\de^\mu_\nu-U^\mu_1(\tau)\,
U_{1\nu}(\tau)\big]\Psi^{\pr\nu}_+(\tau),\label{U1}\\
U^{\pr\mu}_2(\tau)&=&\gamma m_2^{-1}\big[\de^\mu_\nu-U^\mu_2(\tau)\,
U_{2\nu}(\tau)\big]\Psi^{\pr\nu}_-(\tau-\pi),\label{U2}\\
\Psi^{\pr\mu}_-(\tau)&=&\Psi^{\pr\mu}_+(\tau)-
2m_1\gamma^{-1}U^{\pr\mu}_1(\tau),\label{psU1}\\
\Psi^{\pr\mu}_+(\tau+\pi)&=&\Psi^{\pr\mu}_-(\tau-\pi)-
2m_2\gamma^{-1}U^{\pr\mu}_2(\tau).\label{psU2}
\eea
where $\de^\mu_\nu=\left\{\begin{array}{cl} 1, & \mu=\nu\\ 0, & \mu\ne\nu.
\end{array}\right.$
Solving the systems of Eqs.~(\ref{U1}), (\ref{U2}) with the initial conditions
\be
U^\mu_i\big(\check\tau(\lm_i)\big)=v^\mu(\lm_i)\big/\sqrt{v^2(\lm_i)},
\quad i=1,2,
\label{Uini}\ee
we obtain the velocities $U_i^\mu(\tau)$
on the base of the functions $\Psi^{\pr\mu}_\pm(\tau)$ known from
Eqs.~(\ref{psi}) in the initial segments. Simultaneously we continue
(with no limit) the functions $\Psi^\mu_\pm$ outside these segments with
the help of Eqs.~(\ref{psU1}), (\ref{psU2}) and determine the world surface
(\ref{sol}).

The above described procedure of solving the IBVP for the string with
the fixed end
(the infinitely heavy quark with $m_1\to\infty$ is at rest) and $m_2=1$,
$\gamma=1$ is illustrated with Fig.~2. This is a typical example of
a slightly disturbed motion close to the rotational one (\ref{rot}).
In the case $m_1\to\infty$ Eqs.~(\ref{U1}), (\ref{psU1}) are to be
substituted for the equations \cite{BaSh} $U_1^\mu(\tau)=U_1^\mu={}$const
(the infinite mass moves at a constant velocity) and
\be
\Psi^{\pr\mu}_-(\tau)=\big[2 U_1^\mu U_{1\nu}-\de^\mu_\nu \big]\,
\Psi^{\pr\nu}_+(\tau),\qquad m_1\to\infty.
\label{inf}\ee
Eq.~(\ref{U2}) with initial condition (\ref{Uini}), $i=2$ was solved
numerically.

\begin{figure}[bh]   
\unitlength=1mm     
\begin{center}
\begin{picture}(86,8)
\end{picture}\end{center}
\caption{The curved string motion of type (5), (6).}
\end{figure}

For the motion in Fig.~2 the initial distance between the quarks
is $R(0)=1$, the initial string shape (the curve 1 in Figs.~2a and c) is
determined by Eq.~(\ref{fAOV}) with the amplitude multiplier
$\beta_0=\ddot\phi(0)\,\om^{-3}R^{-1}(0)=1$.
The initial velocities of string points correspond to rotating in $xy$-plane
at the angular velocity $\vec\om=\{0;0;\om\}$, $\om=1/\sqrt2$:
\be
v^\mu(\lm)=\{1; \vec v(\lm)\},\qquad
\vec v(\lm)=\big[\vec\om\times\vec\rho(\lm)\big]
\label{vel}\ee
The initial speed of the moving end $|\vec v(\lm_2)|=v_\perp=\om R(0)=1/\sqrt2$
satisfies the relation
\be
v_\perp^2+\om v_\perp m_i/\gamma=1
\label{Rom}\ee
$(i=2)$, that is valid for the rotational motion (\ref{rot}) \cite{Ko,4B}

In Figs.~2a, b the positions of the string in $xy$-plane (sections
$t={}$const of the world surface) are shown. They are numbered in order of
increasing $t$ with the step in time $\Delta t=0.2$ and these numbers are
near the position of the moving quark marked by the small circle.
The first turn of the string is shown in Fig.~2a and the fourth one
(after the missed time interval) is represented in Fig.~2b. We can see
that main features of the motion are kept: the string rotates and its
shape (slightly deviating from straightness) changes quasiperiodically.

The evolution of this shape is shown in details in Fig.~2c, where
the first 9 curves from Fig.~2a (with the same step $\Delta t=0.2$)
 are turned in $xy$-plane so that the 2-nd
end of each curve lies on the axis of abscissas. In other words the curves
in Fig.~2c are the string positions in the frame of reference
rotating with the string.
One may conclude that the shape (\ref{fAOV}) of the string is not conserved
with growing time. Changing the shape looks like spreading waves along
the string (that uniformly rotates). Drawing other curves in this manner
we see that the disturbances of the shape do not grow with growing time $t$.
This also concerns of the distance $R(t)$ between the quarks shown in Fig.~2d.
The distance $R(t)$ vary quasiperiodically near the initial value $R(0)=1$.
It is interesting that the period $T_R\simeq4.75$ of varying $R$ does not
coincide with the rotational period $T=2\pi/\om\simeq8.886$ and the period
$T_s\simeq2$ of changing the string shape. This hierarchy of the periods
or frequencies and the graphs in Fig.~2e will be clarified below.

The picture of motion (including the behavior of $R$ and the string shape)
 is similar for various values of $\om$, $R(0)$, $m_2$
connected by Eq.~(\ref{Rom}), and for various amplitudes $\beta_0$
of the disturbance (\ref{fAOV}). If $\beta_0$ is small enough the string
motion is visually identical with the pure rotational one. But the disturbance
of the string shape does not keep its form. So we may conclude that the
expressions (\ref{AOV}), (\ref{fAOV}) from Ref.~\cite{AllenOV} do not describe
a motion of the curved string during an appreciable time interval.

The problem of searching the quasirotational motions of the curved string,
for which the string shape behaves like stationary waves is solved in this
paper with the more convenient (in comparison with Ref.~\cite{AllenOV})
analytical approach. This method also confirms analytically the numerical
investigation of the stability problem for the rotational motions.

For this purpose we consider the string with the infinitely heavy end
$m_2\to\infty$ (or with the fixed end in the frame of reference moving
at the velocity $U_2^\mu={}$const) for the sake of simplicity.
It is convenient to use the unit velocity vector of the moving end
\be
U_1^\mu(\tau)\equiv U^\mu(\tau),\qquad U^2(\tau)=1
\label{U}\ee
for describing the string motion because the world surface is totally
determined from the given function $U^\mu(\tau)$ and the value
$\gamma/m_1$ with using the formulas
\cite{An,Det}
\be
\Psi^{\pr\mu}_\pm(\tau)=m_1\gamma^{-1}\big[
\sqrt{-U^{\pr2}(\tau)}\,U^\mu(\tau)\pm U^{\pr\mu}(\tau)\big].
\label{det}\ee
If we substitute the analog of Eq.~(\ref{inf}) for the second end
$$\Psi^{\pr\mu}_+(\tau)=P_\nu^\mu\Psi^{\pr\nu}_{(-)},\quad
P_\nu^\mu=2U_2^\mu U_{2\nu}-\de^\mu_\nu,\quad m_2\to\infty$$
and Eq.~(\ref{det}) into Eq.~(\ref{U1}), we get the relation
\be
U^{\pr\mu}(\tau)=(\de^\mu_\nu-U^\mu U_\nu)\,P^\nu_\kappa
\Big[\sqrt{-U^{\pr2}_{(-)}}\,U^\kappa_{(-)}-U^{\pr\kappa}_{(-)}\Big].
\label{cont}\ee
Here and below ${}_{(-)}\equiv(\tau-2\pi)$, the argument $(\tau)$ may
be omitted.

The vector-function $U^\mu(\tau)$ given in a segment with the length $2\pi$
(if $\gamma/m_1$, $U_2^\mu$ are also given) contains all information about
this motion of the system \cite{Det}. Indeed, the system of ordinary
differential equations with shifted argument (\ref{cont}) let
us continue $U^\mu(\tau)$ beyond the given initial segment and obtain
the world surface with using Eqs.~(\ref{det}) and (\ref{sol}).

For the rotational motion (\ref{rot}) the velocity of the moving quark
satisfying Eq.~(\ref{cont}) may be written in the form
\be
\bar U^\mu(\tau)=\Gamma\big[e_0^\mu+v_0\acute e^\mu(\tau)\big],
\qquad\Gamma=(1-v_0^2)^{-1/2}.
\label{Urot}\ee
Here $v_0=v_\perp=\sin\pi\theta$ is the constant speed of this quark;
$\acute e^\mu=\theta^{-1}\frac d{d\tau}e^\mu(\tau)$,
$e^\mu=e_1^\mu\cos\theta\tau+e_2^\mu\sin\theta\tau$ are unit space-like
rotating vectors, $\theta=\om b=\pi^{-1}\arcsin v_0={}$const is the
``frequency" with respect to $\tau$, the ends
are renumerated in comparison with Eq.~(\ref{rot}).
The 2-nd quark in this frame of reference is at rest: $U_2^\mu=e_0^\mu$.

To study the stability of the rotational motion (\ref{rot}) we consider
arbitrary small disturbances of this motion or of the vector (\ref{Urot})
in the form
\be
U^\mu(\tau)=\bar U^\mu(\tau)+u^\mu(\tau).
\label{U+u}\ee
The disturbance $u^\mu(\tau)$ is given in the initial segment,
for example, $I=[-2\pi,0]$; it is to be small $|u^\mu|\ll1$ (we neglect the
second order terms) and satisfies the condition
\be
\big(\bar U(\tau),u(\tau)\big)=0,
\label{Uu}\ee
resulting from Eq.~(\ref{U}) for both $U^\mu$ and $\bar U^\mu$.

Substituting expressions (\ref{U+u}), (\ref{Urot}) into Eq.~(\ref{cont})
we obtain the equation describing the evolution of $u^\mu$
\bea
u^{\pr\mu}&+&Q u^\mu+\bar U^\mu(\bar U'\!\!,u)=
(\de^\mu_\nu-\bar U^\mu\bar U_\nu)\,P^\nu_\kappa
\big[Q u^\kappa_{(-)\!}-u^{\pr\kappa}_{(-)}\big]\nonumber\\
&+&Q^{-1}\big[(1+2v_0^2)\,\bar U^\mu-P^\mu_\nu
\bar U^\nu_{(-)}\big] (\bar U'_{(-)},u'_{(-)}),
\label{uev}\eea
where
$$Q=\sqrt{-\bar U^{\pr2}}=\Gamma\theta v_0=\frac{v_0\arcsin v_0}
{\pi\sqrt{1-v_0^2}}={\mbox{const}}.$$

Projecting the disturbance $u^\mu$ onto the basic vectors $e_0^\mu$,
$e^\mu(\tau)$, $\acute e^\mu(\tau)$, $e_3^\mu$ we denote its three independent
components in the following manner:
$$u_0(\tau)=(e_0,u),\quad u_e(\tau)=(e,u),\quad u_z(\tau)=(e_3,u).$$
The fourth component is expressed due to Eq.~(\ref{Uu}):
$(\acute e,u)=-v_0^{-1}u_0$. Considering corresponding projections
one can transform the system (\ref{uev}) into the following one:
\bea
u'_0+Qu_0-\Gamma Qu_e&=&u'_{0(-)}-Qu_{0(-)}-\Gamma Qu_{e(-)},\label{uev0}\\
u'_e+Qu_e+\theta v_0^{-1}u_0&=&u'_{e(-)}+Qu_{e(-)}+F(u_{0(-)}),
\label{uev1}\\
u'_z+Qu_z&=&u'_{z(-)}-Qu_{z(-)},
\label{uevz}\eea
where $F(u_{0(-)})=\theta(v_0^{-1}+2v_0)\,u_{0(-)}-2\sqrt{1-v_0^2}\,u'_{0(-)}$.

Eq.~(\ref{uevz}) for $z$-axial disturbances is independent on others and
may be easily solved: if $u_z(\tau)$ is given in the initial segment
$I=[-2\pi,0]$, the continuation of this function for $\tau>0$ is
[here $\Delta u_z=u_z(0)-u_z(-2\pi)$]:
\be
u_z(\tau)=u_{z(-)}+e^{-Q\tau}\Big[\Delta u_z-2Q
\int_0^{\tau}e^{Q\tilde\tau}u_{z(-)}\,d\tilde\tau\Big].
\label{uz}\ee

The pure harmonic solutions (\ref{uz}) $u_z(\tau)=e^{i\vartheta\tau}$ exist
if the ``frequency" $\vartheta$ satisfies the relation
\be
\vartheta/Q=\cot\pi\vartheta,
\label{zfreq}\ee
resulting from substitution $u_z=e^{i\vartheta\tau}$ into Eq.~(\ref{uevz}).
The transcendental Eq.~(\ref{zfreq}) has the countable set of roots
$\vartheta_n$, $n-1<\vartheta_n<n$, the minimal positive root
$\vartheta_1=\theta=\pi^{-1}\arcsin v_0$.
These pure harmonic $z$-disturbances corresponding to various $\vartheta_n$
result in the following correction to the motion (\ref{rot}) [due to
Eqs.~(\ref{det}), (\ref{sol}) there is only $z$ or $e_3^\mu$ component of
the correction]:
\be
X^3(\tau,\s)=B\sin\vartheta_n\s\cdot\cos(\vartheta_n\tau+\varphi_0).
\label{zwave}\ee
Here the string ends have been again renumerated so $\s=0$ corresponds
to the fixed end as in Eq.~(\ref{rot}). The amplitude $B$ is to be small
in comparison with $\om^{-1}$.

Expression (\ref{zwave}) describes oscillating string in the form of
orthogonal (with respect to the rotational plane) stationary waves with
$n-1$ nodes in the interval $0<\s<\pi$. Note that the moving quark is not in
a node, it oscillates along $z$-axis at the frequency
$\om_n=\om\vartheta_n/\theta$. The shape $F=B\sin\vartheta_n\s$ of the
$z$-oscillation (\ref{zwave}) is not pure sinusoidal with
respect to the distance $s=\om^{-1}\tilde\s=\om^{-1}\sin\theta\s$ from
the center to a point ``$\s$":
\be
F(s)=B\sin(\vartheta_n\theta^{-1}\arcsin\om s),\quad
0\le s\le v_0/\om.
\label{zshape}\ee
If $n=1$ this dependence is linear. In this trivial case the motion is
pure rotational (\ref{rot}) with a small tilt of the rotational plane.
But the motions (\ref{zwave}) with excited higher harmonics $n=2,3,\dots$
are non-trivial.

Eq.~(\ref{uevz}), its harmonic solutions, the ``frequencies" $\vartheta_n$
as roots of Eq.~(\ref{zfreq}) were investigated in Refs.~\cite{An,PeSh}
where the motions of the meson string with linearizable boundary conditions
(\ref{qq}) were studied and classified. Those solutions
(in $3+1$\,-\,dimensional Minkowski space) with $\vartheta_n$, $n\ge2$
described the exotic motions of the $n-1$ times folded rectilinear string
with $n-1$ points moving at the speed of light. But in Eq.~(\ref{zwave})
these higher harmonics became apparent as (much more physical) excitations of
the rotating string.

It was shown in Ref.~\cite{PeSh} that any smooth function $u(\tau)$
in a segment $I$ with the length $2\pi$ may be expanded in the series
\be
u(\tau)=\sum_{n=-\infty}^{+\infty}u_n\exp(i\vartheta_n\tau),\quad
\tau\in I=[\tau_0,\tau_0+2\pi].
\label{zser}\ee
So if we expand any given disturbance $u_z(\tau)$ in the initial segment $I$,
we obtain the solution (\ref{uz}) of Eq.~(\ref{uevz}) in the form (\ref{zser})
for all $\tau\in R$. If the initial function $u_z(\tau)$ satisfies
Eq.~(\ref{uevz}) at the ends of $I$, $u_z(\tau)\in C^2(I)$, this series
converges absolutely
in $I$, hence its sum (\ref{zser}) is a limited function for all $\tau\in R$.
This is the proof of stability of the rotational motions (\ref{rot}) for
the string with fixed end with respect to the $z$-oscillations.

In Fig.~2 we observe another type of oscillations. These planar disturbances
are described by Eqs.~(\ref{uev0}), (\ref{uev1}). Their solutions
in the form $u_0=\beta_0e^{i\Theta\tau}$, $u_e=\beta_1e^{i\Theta\tau}$
exist only if the ``frequency" $\Theta$ satisfies the equation
\be
\Theta^2-Q^2(1+v_0^{-2})=2Q\Theta\cot\pi\Theta,
\label{pfreq}\ee
The roots of Eq.~(\ref{pfreq}) $\Theta_n\in(n-1,n)$, $n\ge1$
behave like the roots $\vartheta_n$ of Eq.~(\ref{zfreq}) but
$\Theta_n>\vartheta_n$. The disturbances (\ref{U+u})
$u^\mu=\beta^\mu\exp(i\Theta_n\tau)$ correspond to the following small
($B\ll\om^{-1}$) harmonic planar oscillations or planar stationary waves:
\bea
X^\mu&=&e_0^\mu t+e^\mu(\tau)\big[\om^{-1}\sin\theta\s+
B f_r(\s)\cos\Theta_n\tau\big]\nonumber\\
&+&B\big[e_0^\mu f_0(\s)+\acute e^\mu(\tau)\,f_s(\s)\big]\sin\Theta_n\tau,
\label{pwave}\\
f_0&=&2(\theta^2-\Theta_n^2)\cos\Theta_n\s,\;\,
B=\frac{m_i\beta_0}{2\gamma(\Theta_n^2-\theta^2)\sin\pi\Theta_n},
\nonumber\\
f_s&=&(\Theta_n+\theta)^2\sin(\Theta_n-\theta)\s-
(\Theta_n-\theta)^2\sin(\Theta_n+\theta)\s,\nonumber\\
f_r&=&(\Theta_n+\theta)^2\sin(\theta-\Theta_n)\s-
(\Theta_n-\theta)^2\sin(\Theta_n+\theta)\s.\nonumber
\eea
The shape of these stationary waves in the co-rotating frame of reference
(where the axes $x$ and $y$ are directed along $e^\mu$ and $\acute e^\mu$)
is approximately described by the function
$B\big[f_s(\s)-f_0(\s)\sin\theta\s\big]$ if the deflection is maximal.
These shapes $F=F_n(s)$ for $n=1,2,3,4$ are shown in Fig.~2e with indicated
numbers $n$. For each $n$ this curved string oscillates at the frequency
$\om_n=\om\Theta_n/\theta$, it has $n-1$ nodes in $(0,\pi)$ (which are not
strictly fixed because $f_0$ and $f_r$ are non-zero) and the
moving quark is not in a node, especially for the main mode with $n=1$.
The latter feature [similar to the motion
(\ref{zwave})] radically differs from that in expression
(\ref{fAOV}) \cite{AllenOV} where the disturbance in this endpoint
is forcedly nullified. Note that Eq.~(\ref{pwave}) describes both the
deflection of this endpoint $B\acute e^\mu f_s(\pi)\sin\Theta_n\tau$ and
its radial motion $Be^\mu f_r(\pi)\cos\Theta_n\tau$.

Any smooth disturbance $u(\tau)$ may be expanded in the series similar
to Eq.~(\ref{zser}) with the roots $\Theta_n$ of Eq.~(\ref{pfreq}) \cite{PeSh}.
So any quasirotational motion of this sting, in particular, the motion in
Fig.~2 is the superposition of the stationary waves (\ref{pwave}) and
(for nonplanar motions) (\ref{zwave}). This fact let us conclude, that
the rotational motions (\ref{rot}) for the considered model are {\it stable},
and explain the above mentioned problem with the ``hierarchy of the periods"
for the example in Fig.~2. Now, when we expand this solution in the combination
of expressions (\ref{pwave}), it becomes obvious that the radial period $T_R$
(Fig.~2e) corresponds the main ``frequency" $\Theta_1\simeq0.463$ with the
background of higher harmonics, whereas the ``shape" period $T_s$ is
connected mainly with the following  ``frequency" $\Theta_2\simeq1.149$,
because the string shape for the mode $\Theta_1$ is more close to
rectilinear one (Fig.~2e) and it is ``not observable" in Fig.~2c.
Remind that $T_R\Theta_1=T_s\Theta_2=T\theta$ where $T$ is the rotational
period and $\theta=0.25$.

The obtained results are generalized for the string with both finite masses
$0\le m_1,m_2<\infty$ at the ends. For this purpose we are to substitute the
disturbed expression (\ref{U+u}) into Eqs.~(\ref{U1})\,--\,(\ref{psU2}),
(\ref{det}) and deduce the equations generalizing Eqs.~(\ref{uev}).
The search of their oscillatory solutions $u^3=B_z\exp(i\vartheta\tau)$,
 $u^\mu=B^\mu\exp(i\Theta\tau)$ results in the following generalizations
of Eqs.~(\ref{zfreq}) and (\ref{pfreq}):
\bea
(\vartheta-Q_1Q_2/\vartheta)\big/(Q_1+Q_2)&=&\cot\pi\vartheta,\label{gzfreq}\\
\frac{\Phi_1(\Theta)\,\Phi_2(\Theta)-4Q_1Q_2\Theta^2}
{2\Theta\big[Q_1\Phi_2(\Theta)+Q_2\Phi_1(\Theta)\big]}&=&\cot\pi\Theta,
\label{gfreq}\eea
where $Q_i=\theta v_i/\sqrt{1-v_i^2}$, $v_i$ are the constant speeds of the
ends for Eq.~(\ref{rot}), $\Phi_i(\Theta)=\Theta^2-Q_i^2(1+v_i^{-2})$.
Corresponding disturbances of the rotational motion (\ref{rot}) are the
stationary waves behaving similar to Eqs.~(\ref{zwave}), (\ref{pwave}).
In particular, in Eq.~(\ref{zwave}) one should substitute $\sin\vartheta_n\s$
for $\cos(\vartheta_n\s+\phi_n)$. In this case the endpoints also oscillate
at satisfying Eqs.~(\ref{gzfreq}), (\ref{gfreq}) ``frequencies"
$\vartheta_n$ or $\Theta_n$.

All roots of Eqs.~(\ref{gzfreq}), (\ref{gfreq}) are real numbers. Hence,
amplitudes of these harmonics are not increasing (and not decreasing)
and the quasirotational motions constructed as combinations of these
harmonics are {\it stable} (but they are not asymptotically stable).
Note that the minimal positive root of both Eqs.~(\ref{gzfreq}) and
(\ref{gfreq}) equals  $\theta=\om b$. In both cases the corresponding
solutions are trivial: they describe a small tilt of the rotational plane
or a small shift of the rotational center in this plane.

The obtained oscillatory motions may be simulated numerically when we solve
the IBVP with the initial data corresponding to the excitation only one mode.
Such an example for the string with the masses $m_1=2$, $m_2=1$ at the ends
and $\gamma=1$, $\Delta t=0.1$ is represented in Fig.~3: the first 20 positions
of the string are in Fig.~3a and the following ones after 3 turns of the system
are in Fig.~3b. The first quark is marked by the point. Here the initial
position is close to the rectilinear segment with the lengths of its two parts
$R_1\simeq0.179$, $R_2=0.3$ connected with $\om\simeq1.6$ by Eq.~(\ref{Rom})
where $v_\perp=v_{\perp i}=\om R_i$. The corresponding value
$\theta\simeq0.252$ is the first positive root of Eq.~(\ref{gzfreq}).
These roots may be seen in Fig.~3c, where the l.\,h.\,s. of Eq.~(\ref{gzfreq})
is shown as dash line, its analog for Eq.~(\ref{gfreq}) is the blue full line
and the graph of $\cot\pi\Theta$ is the red line.

The initial string position (the curve 1 in Fig.~3a) is the rectilinear
segment but the initial velocities are disturbed in comparison with
Eq.~(\ref{vel}): $\vec v=\big[\vec\om\times\vec\rho\big]+\de\vec v(\lm)$.
The disturbance $\de\vec v$ corresponds to the form of the second planar oscillatory harmonic connected with the root $\Theta_2\simeq1.1277$ of
Eq.~(\ref{gfreq}). The shapes of these harmonics for
$\Theta_1\simeq0.4458$, $\Theta_2$, $\Theta_3\simeq2.0669$,
$\Theta_4\simeq3.045$ with indicated numbers are shown in Fig.~3d.
Fig.~3e illustrates the shapes of the curved string with the step
$\Delta t=0.05$ in the co-rotating (at the frequency $\om$) frame of reference,
the initial shape is the rectilinear red dash line.
In Fig.~3f the deflection $\de(t)$ of the string shape from rectilinear one
(measured in the middle of the string) is shown and also the distance $R(t)$
between the quarks (the red dash line).

Arbitrary slightly disturbed rotational motions may also be obtained
as solutions of the IBVP with certain initial data. Numerical experiments
show their stability and that they are always superpositions of the above
described stationary waves.

The stability problem for the rotational motions for the string baryon models
$q$-$q$-$q$, Y and $\Delta$ (Fig.~1) is more complicated one in comparison
with that for the meson string model. We shall consider all string baryon
configurations in turn starting from the model ``triangle" \cite{Tr,PRTr}.

\begin{figure}[bth]         
\unitlength=1mm
\begin{center}
\begin{picture}(86,8)
\end{picture}\end{center}
\caption{The disturbed rotational motion of the string with
$m_1=2$, $m_2=1$.}
\end{figure}

\section{``Triangle" baryon model}
\label{sectr}

This configuration may be regarded as a closed relativistic string with
the tension $\gamma$ carrying three pointlike masses $m_1$, $m_2$, $m_3$
\cite{Tr,PRTr}. In action (\ref{S}) for the ``triangle" model $N=3$ and
the domain $\Omega$ is divided by the quark trajectories into three domains:
$\Omega=\Omega_0\cup\overline\Omega_1\cup\Omega_2$,
$\Omega_i=\{(\tau,\s):\,\s_i(\tau)<\s<\s_{i+1}(\tau)\}$
(Fig.\,4).
The equations $\s=\s_0(\tau)$ and $\s=\s_3(\tau)$
determine the trajectory of the same (third) quark. It is connected with
the fact that string is closed (so the world surface is tube-like) and may
be written in the following general form \cite {Tr}:
\be
X^\mu(\tau,\s_0(\tau))=X^\mu(\tau^*,\s_3(\tau^*)).\label{cl}\ee
The parameters $\tau$ and $\tau^*$ in these two parametrizations
of just the same line are not equal in general.

The equations of string motion and the boundary conditions at three
quark trajectories result from action (\ref{S}) \cite{Tr,PRTr}.
Derivatives of $X^\mu$ can have discontinuities on the
lines $\s=\s_i(\tau)$ (except for tangential).
However, by choosing coordinates $\tau$, $\s$
the induced metric on all the world surface may be made
continuous and conformally flat \cite {Tr}, i.e.
satisfying the orthonormality conditions (\ref{ort}).
Under these conditions the equations of motion for all
$\Omega_i$ take the form (\ref{eq})
and the boundary conditions are \cite{Tr,PRTr}
\bea
m_i\frac d{d\tau}U_i^\mu(\tau)
&-&\gamma\big[X'{}^\mu+\s_i'(\tau)\,\dot X^\mu\big]
\Big|_{\s=\s_i+0}\nonumber\\
&+&\gamma\big[X'{}^\mu+\s_i'(\tau)\,\dot X^\mu\big]
\Big|_{\s=\s_i-0}\!\!=0,
\label{qqq}\eea
where the same notation $U^\mu_i$
for the unit velocity vector of $i$-th quark is used.
In accordance with Eq.~(\ref{cl}) for the 3-rd quark
one should put $\s=\s_0(\tau)$ in the second term of Eq.~(\ref{qqq})
and replace $\tau$ by $\tau^*$ in the last term.

At this stage we have five undetermined functions in this model:
$\tau^*(\tau)$ in the closure condition (\ref{cl}) and four
trajectories $\s_i(\tau)$, $i=0,\,1,\,2,\,3$.
Using the invariance of Eqs.~(\ref{ort}), (\ref{eq}), (\ref{qqq})
with respect to the reparametrizations (\ref{rep})
we fix two of these functions as follows (Fig.\,4):
\be\s_0(\tau)=0,\qquad\tau^*(\tau)=\tau.
\label{fit}\ee
The remaining three functions will be calculated
with solving the IBVP.

The first Eq. (\ref{fit}) may be obtained
by the substitution (\ref{rep}) with the required $f_+$
and $f_-(\eta)= \eta$. At the second step one can get the equality
$\tau^*=\tau$ by repeating the procedure (\ref{rep}) with the function
$f_+=f_-\equiv f$ that satisfies the condition
$$2f(\tau)=f\big(\tau^*(\tau)+\s_3(\tau)\big)+
f\big(\tau^*(\tau)-\s_3(\tau)\big).$$
The constraint $\tau^*(\tau)=\tau$ (that is the closure of any
coordinate line $\tau={}$const on the world surface) was not fulfilled
in Refs.~\cite{Tr,PRTr}.

\begin{figure}[hb]
\unitlength=1.0mm
\begin{center}
\begin{picture}(84,31)
\put(10,1){\vector(1,0){68}} \put(10,1){\vector(0,1){29.5}}
\put(7.5,0.5){0} \put(77,2){$\s$} \put(11,29){$\tau$}
\put(6,12){$\s_0$} \put(34.5,28){$\s_1(\tau)$}
\put(55.5,28){$\s_2(\tau)$} \put(72,12){$\s_3(\tau)$}
\put(20,20){$\Omega_0$}
\put(41,20){$\Omega_1$} \put(60,20){$\Omega_2$}
\put(19.5,6){$D_0$}\put(40,7){$D_1$}\put(57,6){$D_2$}
\put(10,1){\line(1,1){22.75}} \put(69,1){\line(-1,1){15.75}}
\put(31.5,2.5){\line(1,1){23.5}} \put(31.5,2.5){\line(-1,1){21.5}}
\put(51,3.5){\line(-1,1){19}} \put(51,3.5){\line(1,1){23}}
\thicklines
\put(10,1){\line(0,1){27}}
\put(10,1){\line(6,-1){6}} \put(16,0){\line(1,0){2}}
\put(18,0){\line(5,1){7.5}} \put(25.5,1.5){\line(6,1){18}}
\put(43.5,4.5){\line(1,0){2.5}}
\put(46,4.5){\line(5,-1){7.5}} \put(53.5,3){\line(6,-1){6}}
\put(59.5,2){\line(1,0){3.5}} \put(63,2){\line(6,-1){6}}
\put(31.5,2.5){\line(-1,6){1}} \put(30.5,8.5){\line(0,1){3.5}}
\put(30.5,12){\line(1,6){1.5}} \put(32,21){\line(1,5){1}}
\put(33,26){\line(1,4){1}}
\put(51,3.5){\line(0,1){2}} \put(51,5.5){\line(1,6){1}}
\put(52,11.5){\line(1,5){1}} \put(53,16.5){\line(1,4){1}}
\put(54,20.5){\line(1,6){1}} \put(55,26.5){\line(0,1){3.5}}

\put(69,1){\line(1,6){1}} \put(70,7){\line(1,5){1}}
\put(71,12){\line(1,4){1}} \put(72,16){\line(1,5){1}}
\put(73,21){\line(1,6){1.5}}
\end{picture}
\caption{The domain $\Omega$ for the model $\Delta$.}
\label{figOm}
\end{center}
\end{figure}
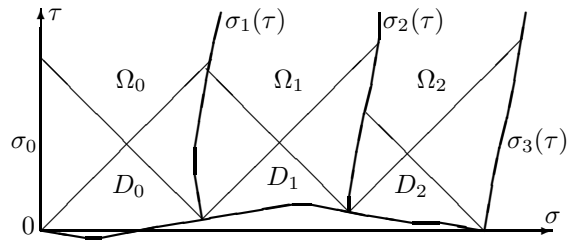

Because of discontinuities of $X^{\pr\mu}$ at
$\s=\s_i(\tau)$ the general solutions of Eq.~(\ref{eq})
in three domains $\Omega_i$ are described by three different functions
($i=0,1,2$)
\be
X^\mu(\tau,\s)=\frac1{2}\bigl[\Psi^\mu_{i+}(\tau+\s)+\Psi^\mu_{i-}(\tau-\s)
\bigr],\quad (\tau,\s)\in\Omega_i.
\label{solt}\ee
Nevertheless, the function $X^\mu$ (with the tangential derivatives)
is continuous in $\Omega$.

The initial-boundary value problem (IBVP) for the ``triangle" string
configuration is stated similarly to that for the meson model in
Sect.~\ref{sec1}. The procedure of its solving is briefly described
below. One can find the details in e-print \cite{sttr}.

An initial position of the string can be given as the curve (\ref{curve})
in Minkowski space but for the $\Delta$-model $\lm\in[\lm_0,\lm_3]$
and this curve is closed: $\rho^\mu(\lm_0)=\rho^\mu(\lm_3)$.
The function $\rho^\mu(\lm)$ is piecewise smooth, $\rho^{\pr\mu}$ may have
discontinuities at the quark positions $\lm=\lm_1,\,\lm_2$.

Initial velocities on the initial curve can be given as a time-like
vector $\,v^\mu(\lm)$, $\lm\in[\lm_0,\lm_3]$, $v^\mu(\lm)$ may be
multiplied by an arbitrary scalar function $\phi(\lm)>0$.
The condition $v^\mu(\lm_0)=v^\mu(\lm_3)\cdot{}$const is fulfilled.

To solve the problem we use the parametrization (\ref{tasi}), (\ref{X=r}),
$\lm\in[\lm_0,\lm_3]$ of the initial curve on the world surface
(Fig.~4) and the formulas (\ref{psi}) for determining
the functions $\Psi^{\pr\mu}_{i\pm}$
from the initial data in the finite segments
\bea
\Psi^\mu_{i+}(\xi),&&\quad\xi\in\bigl[\check\tau(\lm_i)+\check\s(\lm_i),
\check\tau(\lm_{i+1})+\check\s(\lm_{i+1})\bigr],\nonumber\\
\Psi^\mu_{i-}(\xi),&&\quad\xi\in\bigl[\check\tau(\lm_{i+1})-
\check\s(\lm_{i+1}),\check\tau(\lm_i)-\check\s(\lm_i)\bigr],
\label{pseg}\eea
that lets us find the solution of the problem in the form (\ref{solt})
in the zones $D_i$ shown in Fig.~4.
In these zones bounded by the initial curve and the characteristic lines
$\tau-\s\leq\check\tau(\lm_i)-\check\s(\lm_i)$,
$\tau+\s\leq\check\tau(\lm_{i+1})+\check\s(\lm_{i+1})$
the solution depends only on initial data
without influence of the boundaries.
In others parts of the domains $\Omega_i$ the solution is obtained
with the help of the boundary conditions (\ref{qqq}) which may be reduced
to the form \cite{sttr}
\bea
U^{\pr\mu}_i&=&\gamma m_i^{-1}\big[\de^\mu_\nu-U^\mu_i(\tau)\,
U_{i\nu}(\tau)\big]\nonumber\\
&\times&\frac d{d\tau}\big[\Psi^\nu_{i+}(\tau+\s_i)+
\Psi^\nu_{(i-1)-}(\tau-\s_i)\big].
\label{Uqi}\eea
For $i=3$ in Eq.~(\ref{Uqi}) one should replace
$\Psi^\mu_{3+}(\tau+\s_3)$ by $\Psi^\mu_{0+}(\tau)$ in accordance with
Eqs.~(\ref{cl}) and (\ref{fit}).

Integrating systems (\ref{Uqi}) with the initial conditions (\ref{Uini}),
$i=1,2,3$ we can determine unknown vector functions $U^\mu_i(\tau)$ for
$\tau\in[\check\tau(\lm_i),\tau_i^c]$
with the help of the functions $\Psi^{\pr\mu}_{i\pm}$ known in the segments
(\ref{pseg}) from the initial data. Here $\tau_i^c$ are (minimal) ordinates
of the points in which the trajectories $\s=\s_i(\tau)$
cross the characteristic lines $\tau\pm\s={}$const (Fig.~4).
However we can continue this procedure for $\tau>\tau_i^c$ if for every value
of $\tau$ [after calculating $U_i^{\pr\mu}$ from Eq.~(\ref{Uqi})] we determine
$\Psi^{\pr\mu}_{i\pm}$ outside segments (\ref{pseg}) from the equations
($i=1,2$)

\bea
\Psi^\mu_{(i-1)+}(\tau+\s_i)=\Psi^\mu_{i+}(\tau+\s_i)&-
\frac{m_i}\gamma U^\mu_i(\tau)+&C_i^+,\nonumber\\
\Psi^\mu_{i-}(\tau-\s_i)=\Psi^\mu_{(i-1)-}(\tau-\s_i)&-
\frac{m_i}\gamma U^\mu_i(\tau)+&C_i^-.
\label{psipr}\eea
They are obtained from Eqs.~(\ref{qqq}), (\ref{solt}) and from
continuity of $X^\mu$ \cite{sttr}. The similar relations for $i=3$ are
\begin{eqnarray*}
&\Psi^\mu_{2+}(\tau+\s_3)=\Psi^\mu_{0+}(\tau)
-m_3\gamma^{-1}U^\mu_3(\tau)+C_3^+,&\\
&\Psi^\mu_{0-}(\tau)=\Psi^\mu_{2-}(\tau-\s_3)-
m_3\gamma^{-1}U^\mu_3(\tau)+C_3^-.&
\end{eqnarray*}
The constants of integration $C^\pm_i$ are fixed from Eq.~(\ref{X=r}).

For solving the system (\ref{Uqi}) we are to determine the functions
$\s_i(\tau)$ for $\tau>\check\tau(\lm_i)$. Multiplying $U^\mu_i(\tau)$ by
$\frac d{d\tau}\Psi^\mu_{j\pm}(\tau\pm\s_i)$ we obtain the equalities
$$(1-\s'_i)\big(U_i,\Psi'_{(i-1)-}(\tau-\s_i)\big)
=(1+\s'_i)\big(U_i,\Psi'_{i+}(\tau+\s_i)\big)$$
which let us express $\s'_i$ in the following way (separately for $i=1,2$ and
$i=3$):
\bea
\frac{d\s_i}{d\tau}&=&
\frac{\big(U_i,[\Psi'_{(i-1)-}(\tau-\s_i)-\Psi'_{i+}(\tau+\s_i)]\big)}
{\big(U_i,[\Psi'_{(i-1)-}( \tau-\s_i)+\Psi'_{i+}(\tau+\s_i)]\big)},
\label{sigma}\\
\frac{d\s_3}{d\tau}&=&1-\frac{\big(U_3,\Psi'_{0+}(\tau)\big)}
{\big(U_3,\Psi'_{2-}(\tau-\s_3)\big)}.\nonumber
\eea
Here $U^\mu_i\equiv U^\mu_i(\tau)$.

Eqs.~(\ref{Uqi})\,--\,(\ref{sigma}) allow us to continue the functions
$\Psi^\mu_{i\pm}$ unambiguously beyond the segments (\ref{pseg}).
This algorithm solves (numerically in general) the considered IBVP
with arbitrary initial conditions $\rho^\mu(\lm)$, $v^\mu(\lm)$.

For the model ``triangle" the exact solutions of Eqs.~(\ref{eq}), (\ref{ort}),
(\ref{cl}), (\ref{qqq}) describing rotational motions of the system
(with hypocycloidal
segments of the strings between the quarks) are obtained in
Refs.~\cite{Tr,PRTr}. World surfaces with the parametrization
$\s_i(\tau)=\s_i={}$const and the form of closure condition (\ref{cl})
$\tau^*-\tau=T={}$const may be represented as follows
($X^1\equiv x$, $X^2\equiv y$):
\be X^0=\tau-\alpha\s,\qquad X^1+iX^2=w(\s)\,e^{i\om\tau}.
\label{hyp}\ee
Here $w(\s)=A_i\cos\om\s+B_i\sin\om\s$, $\s\in[\s_i,\s_{i+1}]$;
$\alpha=T/(\s_3-\s_0)={}$const, $\om$ is the angular frequency of this rotation.
Real ($\s_i$, $T$, $\om$, $m_i/\gamma$) and complex ($A_i$, $B_i$) constants
are connected by certain relations \cite{Tr,PRTr,InSh}. There are many
topologically different types\footnote{The so called exotic states
\cite{PRTr} contain string points moving at the speed of light.}
of solutions (\ref{hyp}) but we shall consider motions close to the simple
states \cite{PRTr} in which three quarks are connected by smooth segments
of hypocycloids.

The simple motions (\ref{hyp}) may be obtained by solving the IBVP with
corresponding initial data. To research the stability of these motions we
consider here the disturbed initial conditions $\rho^\mu(\lm)$ and
$v^\mu(\lm)$. With an illustrative view in Fig.~5 the example of such a motion
[close to the simple state (\ref{hyp})] of the ``triangle" configuration
is represented. Here the initial string position is the rectilinear
equilateral triangle with the base $a=0.5$ and the altitude $h=0.33$ in the
$xy$-plain (position 1 in Fig.~5a). The initial velocities
$v^\mu(\lm)$ in the form (\ref{vel}) correspond to the uniform rotation of the
system at the frequency $\vec\om=\{0;0;\om\}$, $\om=2$ about the origin of
coordinates. The quarks with masses $m_1=1$, $m_2=1.5$, $m_3=1.2$, $\gamma=1$
are placed at the corners of the triangle.

The results of computing are represented as projections of the world surface
level lines $t=X^0(\tau,\s)={}$const onto the $xy$ plain as in Figs~2,\,3
and numbered with the time step $\Delta t=0.15$.
The values $\rho^\mu(\lm)$ and $v^\mu(\lm)$ are close to those
which give the exact hypocycloidal solution (\ref{hyp}) for this system (with
the same $m_i$, $\gamma$ and $\om=2$) describing the uniform rotation of the
system with the string shape that is x-marked in Fig.~5a. The positions of the third quark with $m_3=1.2$ are marked by circles.

The further evolution of the system after one turn of the triangle is
represented in Fig.~5b (one can find the omitted phases of this motion in
Ref.~\cite{sttr}). In Fig.~5c the dependence of three mutual distances
$R_{ij}$ between the three quarks on time $t$ is shown, in particular,
$R_{12}(t)$ is denoted by the full line. In Fig.~5d the deflection $\de(t)$
from rectilinearity of the string segment with $R_{13}(t)$ is shown.
We see that when the system rotates the distances between the quarks and the
configuration of the string segments fluctuate near the
values corresponding to the motion (\ref{hyp}) (x-marks in Fig.~5a).

This situation is typical for slightly disturbed rotational motions of
the  string baryon model ``triangle". Numerous tests (with various values
$m_i$, the energy, $\rho^\mu(\lm)$, $v^\mu(\lm)$ and various types of
disturbances \cite{sttr}) show that the simple rotational motions (\ref{hyp})
are {\it stable}. That is small disturbances of the motions (initial
conditions) do not grow with growing time.

Emphasize that the simple motions (\ref{hyp}) are stable with
respect to transforming into the ``quark-diquark" states of the $\Delta$
string configuration \cite {PRTr} with merging two quarks into the diquark.
It is shown in Ref.~\cite{sttr} that such a transformation may be obtained only
through very strong disturbances of the initial conditions,
for example, by essential reducing one of the sides of the initial triangle.
However, in this case the two nearest quarks do not merge but revolve with
respect to each other.

When one of the quark masses $m_i$ is larger than the sum of two others the  shape of the triangle configuration for the simple rotational motion (\ref{hyp})
tends to a rectilinear segment with the position of the heaviest mass
at the rotational center when the energy of the system decreases \cite{PRTr}.
In other words the $\Delta$ configuration tends to the $q$-$q$-$q$ one.
But the motion of the ``triangle" model remains stable unlike that of the
$q$-$q$-$q$ model described in the next section.

\section{Linear string configuration}
\label{li}

The dynamics of the linear ($q$-$q$-$q$) string baryon model is described by
action (\ref{S}) where $N=3$ and the domain $\Omega$ is divided by the middle quark trajectory into two parts $\Omega_1$ and $\Omega_2$:
$\Omega_i=\{(\tau,\s):\,\s_i(\tau)<\s<\s_{i+1}(\tau)\}$,
$\s_1(\tau)<\s_2(\tau)<\s_3(\tau)$ \cite{lin}.
The equations of string motion under conditions (\ref{ort}) may be
reduced to the same form (\ref{eq}). The boundary conditions at the ends
$\s=\s_1$ and $\s=\s_3$ look like Eqs.~(\ref{qq}) but for the middle
quark they take the form (\ref{qqq}), $i=2$. At this line
the derivatives of $X^\mu$ are not continuous in general.

Using the reparametrizations (\ref{rep}) we fix two (from three)
functions $\s_i(\tau)$ in the form similar to (\ref{fiq}):
$\s_1=0$, $\s_3=\pi$. The third function $\s_2(\tau)$ is obtained from
Eq.~(\ref{sigma}) with solving IBVP that is described in details in
in Ref.~\cite{lin}. The main stages of this procedure are the same as in
Sects.~\ref{sec1}, \ref{sectr}: determining the functions $\Psi^\mu_{i\pm}$,
$i=1,2$
in Eq.~(\ref{solt}) with the help of given $\rho^\mu(\lm)$, $v^\mu(\lm)$ and
Eq.~(\ref{psi}) in the initial segments (\ref{pseg}). The further continuation
of $\Psi^\mu_{i\pm}$ beyond (\ref{pseg}) includes the equations similar to
(\ref{U1})\,-\,(\ref{psU2}) for the endpoints (here $i=3$ for the second end)
and Eqs.~(\ref{Uqi}), (\ref{psipr}) for the middle quark.

The rotational motion of the $q$-$q$-$q$ system is described by Eq.~(\ref{rot})
but with the middle quark at the center of rotation. The authors of
Ref.~\cite{Ko} assumed that motion is unstable with respect to centrifugal
moving away of the middle quark and transforming this configuration into the
quark-diquark one. The numerical experiments were made in Ref.~\cite{lin} in
accordance with the above scheme of solving the IBVP. They showed that the
rotational motion of the $q$-$q$-$q$ system is {\it unstable} indeed.
Any arbitrarily small disturbances of the initial data result in the
complicated motion of the middle quark including its centrifugal moving away
but the material points never merge and the configuration never transforms
into $q$-$qq$ one on the classic level.

In Fig.~6 the example of such a motion of the $q$-$q$-$q$ system is represented
(Figs.~6a\,--\,6d) in comparison with the similar motion of the model
``triangle" (Figs.~6e\,--\,6g).
For both models the quark masses are $m_1=m_3=1$, $m_2=3$, $\Delta t=0.15$
the tension is $\gamma_\Delta=1$ in the ``triangle" and
$\gamma=2$ in the $q$-$q$-$q$ configuration.
The initial shape of the string is the rectilinear segment, for the $\Delta$
configuration it is a particular case of the hypocycloidal motion (\ref{hyp}).
The initial velocities satisfies Eqs.~(\ref{vel}), (\ref{Rom}) where
$v_\perp=v_0=0.5$.
The position of the middle quark (marked by the square) is slightly
displaced with respect to the center of rotation so it uniformly moves at the
initial stage (Figs.~6a, 6e) where the behavior if both systems practically
coincides.

Further (starting with position 17 in Fig.~6b) one can see that in the
$q$-$q$-$q$ model the middle heavy quark moves to the string end, while in the
``triangle" model (Figs.~6f, 6g) it remains in the vicinity of the rotational
center. The latter configuration is stable unlike the $q$-$q$-$q$ one.
The axes are omitted here for saving in space.

But the minimal distance between the nearest two quarks for the $q$-$q$-$q$
system never equals zero. The middle quark begins to play a role of rotational center for this string segment (Fig.~6c) and then it returns to the center of
the system (Fig.~6d) and the process recurs. Such a quasiperiodic motion of
the $q$-$q$-$q$ system is the qualitatively universal result of the evolution
for motions (\ref{rot}) with various types of disturbances \cite{lin}.

\section{Three-string model}
\label{Y}

In the three-string baryon model or Y-configuration
\cite{AY,PY,Collins,Koshkar}
three world sheets (swept up by three segments of the relativistic string)
are parametrized with three different functions $X_i^\mu(\tau_i,\s)$.
It is convenient to use the different notations
$\tau_1$, $\tau_2$, $\tau_3$ for the ``time-like" parameters \cite{Collins,sty}.
But the ``space-like" parameters are denoted here by the same symbol $\s$.
These three world sheets are joined along the world line of the junction
that may be set as $\s=0$ for all sheets without loss of generality
(see below and Ref.~\cite{sty}).

Under these notations the action of the three-string is close to (\ref{S}):
\be
S=-\sum_{i=1}^3\int d\tau_i\bigg[\gamma\!\!\!\int\limits_0^{\s_i(\tau_i)}
\!\!\!\sqrt{-g_i}\,d\s+m_i\sqrt{\dot x_i^2(\tau_i)}\,\bigg].
\label{SY}\ee
This action with different $\tau_i$ generalizes the similar
expressions in Refs.~\cite{AY,PY,Collins} (where $m_i=0$)
and in Ref.~\cite{Koshkar}, where $m_i\ne0$ but the class of motions is
limited.

There is the additional boundary condition at the junction taking the form
\be
X_1^\mu\big(\tau,0\big)=X_2^\mu\big(\tau_2(\tau),0\big)=
X_3^\mu\big(\tau_3(\tau),0\big),
\label{junc}\ee
if the parameters $\tau_i$ on the three world sheets are connected at this line
in the following manner:
$$\tau_2=\tau_2(\tau),\quad\tau_3=\tau_3(\tau),\quad\tau_1\equiv\tau.$$

The equations of motion (\ref{eqn}) and the boundary conditions for the
junction and the quark trajectories are deduced from action (\ref{SY}).
Using its invariance one may choose the coordinates in which the orthonormality
conditions (\ref{ort}) are satisfied and string equations of motion for all
$X_i^\mu$ take the form (\ref{eq}). The junction condition (\ref{junc})
unlike more rigid condition with $\tau_1=\tau_2=\tau_3$ on the junction line
\cite{AY,PY} let us choose these coordinates independently on each world sheet.
After this substitution (new coordinates are also denoted $\tau_i,\s$)
the inner equations of the junction line will be more general
$\s=\s_{0i}(\tau_i)$ (in comparison with the previous ones $\s=0$)
and the boundary conditions in the junction and on the quark
trajectories $\s=\s_i(\tau_i)$ will take the following form:
\bea
\sum_{i=1}^3\big[X_i^{\pr\mu}(\tau_i,\s_{0i})
+\s_{0i}'(\tau_i)\,\dot X_i^\mu(\tau_i,\s_{0i})\big]
\tau'_i(\tau)&=&0,\label{qy}\\
m_iU_i^{\pr\mu}(\tau_i)+\gamma \big[X_i^{\pr\mu}+\s_i'(\tau_i)
\,\dot X_i^\mu\big]\Big|_{\s=\s_i(\tau_i)}&=&0.
\label{qqy}\eea
Here in Eq.~(\ref{qy}) $\tau_i=\tau_i(\tau)$,
$\s_{0i}=\s_{0i}\big(\tau_i(\tau)\big)$.

The reparametrizations similar to (\ref{rep})
\be
\tilde\tau_i\pm\tilde\s=f_{i\pm}(\tau_i\pm\s),\qquad i=1,2,3
\label{repy}\ee
(with six arbitrary smooth monotone functions $f_{i\pm}$) keep
invariance of Eqs.~(\ref{ort}), (\ref{eq}), (\ref{qy}), (\ref{qqy}).
Choosing the functions $f_{i\pm}$
we can fix the equations of the junction and quark trajectories
on each world sheet independently in the form (\ref{fiq})
\be
\s_{0i}(\tau_i)=0,\quad\s_i(\tau)=\pi,\quad i=1,2,3,
\label{fiqy}\ee
One can obtain the first Eq.~(\ref{fiqy}) like that in Eq.~(\ref{fit})
and the equalities $\s_i=\pi$ through the substitution
(\ref{repy}) with $f_{i+}=f_{i-}$ keeping invariance of the equation $\s=0$.

In this paper the parametrization satisfies the conditions
(\ref{fiqy}) and (\ref{ort}). But the ``time parameters"
$\tau_1$, $\tau_2$ and $\tau_3$ in Eq.~(\ref{junc}) are not equal
in general. The possible alternative approach
implies introducing the condition $\tau_2(\tau)=\tau_3(\tau)=\tau$
on the junction line (\ref{junc}) in conjunction with
the condition $\s_{0i}=0$ (or $\s_{0i}={}$const).
But under these restrictions two of the functions $\s_i(\tau)$ on the quark
trajectories are not equal to constants in general.

If under orthonormal gauge (\ref{ort}) we demand satisfying
as conditions (\ref{fiqy}), as the equalities $\tau_1=\tau_2=\tau_3$
on the junction line (\ref{junc}) (as, for example, in Ref.~\cite{PY}),
then we actually restrict the
class of motions of the system, which the model describes.
In other words, not all physically possible motions satisfy
the above mentioned conditions.

The proof of these statements in Ref.~\cite{sty} uses the fact that under
restrictions (\ref{ort}) only reparametrizations (\ref{repy}) with the
functions $f_{i+}=f_{i-}=f_i$ satisfying the condition (\ref{MP})
keep Eqs.~(\ref{fiqy}) \cite{sty,PeSh}.
These functions have the properties: if $f(\xi)$ and $g(\xi)$ satisfy the
conditions (\ref{MP}), then the inverse function $f^{-1}(\xi)$ and the
superposition $f\big(g(\xi)\big)$ also satisfy (\ref{MP}).
To obtain the equalities $\tilde\tau_2=\tilde\tau_3=\tilde\tau_1$
on the junction line $\s=0$, we have to use transformations (\ref{repy}),
(\ref{MP}) satisfying the relations $\tau_i(\tau)=f_i^{-1}\big(f_1(\tau)\big)$,
$i=2,3$. This is possible only if the functions $\tau_2(\tau)$,
$\tau_3(\tau)$ satisfy conditions (\ref{MP}), which are not fulfilled
for an arbitrary motion in general \cite{sty}.

For describing an arbitrary motion of the three-string in the suggested
approach the unknown functions $\tau_i(\tau)$ are determined
from dynamic equations with solving the IBVP for this system.
This approach is similar to the above-formulated one for other string models.
In particular, for the Y configuration we have the general solution
(\ref{solt}) where the index $i$ in $\Psi^\mu_{i\pm}$ numerates the world
sheets. Using the given initial position of the three-string in the form of
three joined curves in Minkowski space
$$x^\mu=\rho^\mu_i(\lm),\quad\lm\in[0,\lm_i],\quad
\rho^\mu_1(0)=\rho^\mu_2(0)=\rho^\mu_3(0)$$
and initial velocities $v_i^\mu(\lm)$ we obtain the functions $\Psi^\mu_{i+}$
and $\Psi^\mu_{i-}$ in the initial segments
$\bigl[0,\check\tau_i(\lm_i)+\pi\bigr]$ and
$\bigl[\check\tau_i(\lm_i)-\pi,0\bigr]$ correspondingly from
Eq.~(\ref{psi}).

The functions $\Psi^\mu_{i\pm}$ are to be continued
with using the boundary conditions (\ref{junc}), (\ref{qy}), (\ref{qqy}).
In particular, the conditions on the quark trajectories (\ref{qqy})
are reduced to the form (\ref{U2}), (\ref{psU2}), (\ref{Uini}):
\bea
&U^{\pr\mu}_i=\gamma m_i^{-1}\big[\de^\mu_\nu-U^\mu_i(\tau_i)\,
U_{i\nu}(\tau_i)\big]\Psi^{\pr\nu}_{i-}(\tau_i-\pi),&\label{Uy}\\
&\Psi^{\pr\mu}_{i+}(\tau_i+\pi)=\Psi^{\pr\mu}_{i-}(\tau_i-\pi)-
2m_i\gamma^{-1}U^{\pr\mu}_i(\tau_i).&\label{psy}
\eea

Substituting Eq.~(\ref{solt}) for $X_i^\mu$ into the boundary conditions in
the junction (\ref{junc}), (\ref{qy}), (\ref{fiqy}) we express the function
$\Psi^{\pr\mu}_{i-}(\tau_i)$ through $\Psi^{\pr\mu}_{i+}(\tau_i)$:
\be
\frac d{d\tau}\Psi^\mu_{i-}\big(\tau_i(\tau)\big)=\sum_{j=1}^3T_{ij}
\frac d{d\tau}\Psi^\mu_{j+}\big(\tau_j(\tau)\big),
\label{jps}\ee
where $T_{ij}=\left\{\begin{array}{rc} -1/3, & i=j\\
2/3, & i\ne j \end{array}\right.$.

Eqs.~(\ref{jps}), (\ref{psy}) and (\ref{Uy}) let us infinitely continue
the functions $\Psi^\mu_{i\pm}$ outside the initial segments if the
functions $\tau_2(\tau)$ and $\tau_3(\tau)$ are known. They can be found with
using the isotropy condition $\Psi^{\pr2}_{i\pm}(\tau_i)=0$ [resulting from
Eqs.~(\ref{ort})] and the following consequences of Eqs.~(\ref{junc}):
$$
[\tau'_i(\tau)]^2\big(\Psi'_{i+}(\tau_i),\Psi'_{i-}(\tau_i)\big)=
\big(\Psi'_{1+}(\tau),\Psi'_{1-}(\tau)\big).
$$
Substituting $\Psi^{\pr\mu}_{i-}$ from Eq.~(\ref{jps}) into these relations
we obtain the formulas for calculating the functions $\tau_i(\tau)$
\be
\tau'_2(\tau)=\frac{(\Psi'_{1+},\Psi'_{3+})}{(\Psi'_{2+},\Psi'_{3+})},\quad
\tau'_3(\tau)=\frac{(\Psi'_{1+},\Psi'_{2+})}{(\Psi'_{2+},\Psi'_{3+})}.
\label{taus}\ee
Here the functions $\Psi^{\pr\mu}_{i+}\equiv\Psi^{\pr\mu}_{i+}(\tau_i)$
are taken from Eqs.~(\ref{psi}) and (\ref{psy}) during solving the IBVP.

Eqs.~(\ref{Uy})\,--\,(\ref{taus}) form the closed system for infinite
continuation of $\Psi^\mu_{i\pm}$ or for solving the IBVP in the three-string
model. The described method is used here (and with more details in
Ref.~\cite{sty}) for investigating the rotational stability of this
configuration. For this purpose we consider the IBVP with disturbed initial
conditions $\rho_i^\mu(\lm)$ and $v_i^\mu(\lm)$.

As was mentioned above the rotational motion of the three-string is uniform
rotating of three rectilinear string segments joined
in a plane at the angles 120${}^\circ$ \cite{AY,PY}. Their lengths $R_i$ or
the speeds $v_i=\om R_i$ are connected with the angular velocity $\om$
by the relation (\ref{Rom}) or
\be
R_i\om^2(R_i+m_i/\gamma)=1.
\label{Romy}\ee
This motion and slightly disturbed motions may be obtained
(numerically, in general) by solving the IBVP
with appropriate initial position $\rho_i^\mu(\lm)$ in the form
three rectilinear segments with lengths $R_i$ and velocities (\ref{vel}) with
some disturbances $\de\rho_i^\mu(\lm)$ or $\de v_i^\mu(\lm)$.

The typical example of a quasirotational motion of the
three-string with masses $m_1=1$, $m_2=2$, $m_3=3$, $\gamma=1$ is represented
in Fig.~7. Here the positions of the system in $xy$-plane are numbered in order
of increasing $t$ with the step in time $\Delta t=0.125$ and these numbers are
near the position of the first quark marked by the small square.
This motion is close to the rotational one: the initial velocities satisfy the
relation (\ref{vel}), $\de v_i^\mu=0$, the angular velocity $\om\simeq1.6$
and the different lengths $R_1=0.3$, $R_3\simeq0.125$ are connected by
Eqs.~(\ref{Romy}). But the assigned value $R_2=0.22$ does not satisfy
(\ref{Romy}) (that gives $R_2\simeq0.179$) so this difference plays a role of
the disturbance for the motion in Fig.~7.

The evolution of this disturbance includes the motion of the junction
(Fig.~7a, b) with varying the lengths of the string segments unless one of
these lengths becomes equal to zero, i.e. the third quark falls into the
junction merges with the junction after the shown position 31 in Fig.~7c.
They move together (Fig.~7d) during the finite time. The waves from the point of
merging spread along the strings and complicate the picture of motion.
Falling $i$-th material point into the junction is simultaneous with the
becoming infinite the corresponding ``time" ($\tau_i\to\infty$). This is not
``bad parametrization" but the geometry of the system changes:
the three-string transforms into the $q$-$q$-$q$ configuration
after merging a quark with the junction. The lifetime of this ``$q$-$q$-$q$
stage" is finite but non-zero because the material point with the mass $m_i$
moving at a speed
$v<1$ can not slip through the junction instantaneously. Otherwise under three
non-compensated tension forces the massless junction will move
at the speed of light. But we must note, that this description is pure
classical one: it will unlikely be the same after developing a more general
QCD-based theory.

Nevertheless on the classic level the numerical experiments in Ref.~\cite{sty}
show that the picture of motion in Fig.~7 is qualitatively identical for any
small asymmetric disturbance $\de\rho_i^\mu(\lm)$ or $\de v_i^\mu(\lm)$.
Starting from some point in time the junction begins to move.
During this complicated motion the distance between the junction and
the rotational center increases and the lengths of the string segments vary
quasiperiodically unless one of the material points inevitably merges with the junction.
So one may conclude that rotational motions of the three-string are
{\it unstable}. The evolution of the instability is slow at the first stage
if the disturbance is small, but the middle and final stages are rather similar
to the motion in Fig.~7b, c.
The dependencies $\tau_i(\tau)$ for these motion do not satisfy the periodicity
conditions (\ref{MP}) in general \cite{sty}. This fact does not allow describing these motions in the frameworks of the parametrization \cite{AY,PY} with
$\tau_1=\tau_2=\tau_3$.
The above-described behavior of slightly disturbed rotational motions
takes place also for the massless ($m_i=0$) three-string model \cite{sty}.

\section*{Conclusion}

In this paper the classic motions of the various meson and baryon (Fig.~1)
string models close to the rotational motions (\ref{rot}) or (\ref{hyp})
are investigated. For the meson string model (or the $q$-$qq$ baryon
configuration) it is made both analytically and numerically but in each
of these methods we use the orthonormal conditions (\ref{ort}) which
let us reduce the problem to solving the systems of ordinary differential
equations (\ref{U1}), (\ref{U2}), (\ref{cont}). Using this approach
we obtained a set of solutions (\ref{zwave}), (\ref{pwave}) describing
small oscillatory excitations of the rotating string in the form of
stationary waves. They are divided into two classes: the orthogonal or
$z$-oscillations (\ref{zwave}) and the planar oscillations (\ref{pwave}).
Each class contains the countable set of solutions with different
``frequencies" $\vartheta_n$, $\Theta_n$, which are the roots of
Eqs.~(\ref{zfreq}), (\ref{pfreq}), (\ref{gzfreq}), (\ref{gfreq}).
For these stationary waves the moving quarks are not in a node of oscillation,
they also oscillate. This was one of the reasons resulting
in the wrong expression (\ref{fAOV}) in Ref.~\cite{AllenOV}.

The energy $M$ and the angular momentum $J$ of the oscillatory excited motions
(\ref{zwave}), (\ref{pwave}) are close to the values $M$ and $J$ \cite{4B}
for the pure rotational motions (\ref{rot}) because the disturbances are
small. But these states (precisely speaking, their generalizations) in the case
of strong disturbances may be used for describing the meson states, which are interpreted as higher radially excited states in the potential models
\cite{InSh}. Note that the planar oscillations (\ref{pwave}) include also
the radial motions of the string endpoints (a quark trajectory in the
co-rotating frame of reference is an ellipse). This especially concerns of
the mode with the ``frequency" $\Theta_1$ of Eq.~(\ref{pfreq}) or
(\ref{gfreq}).

A search of quasirotaional motions in the form similar to (\ref{U+u}) for the
string baryon model ``triangle", $q$-$q$-$q$ and Y encounters essential
difficulties connected with non-fixed quark trajectories $\s=\s_i(\tau)$
in Eq.~(\ref{qqq}) or the expressions $\tau_i(\tau)$ in Eq.~(\ref{junc})
for the three-string configuration.
But for each string baryon model the method of solving the initial-boundary
value problem with arbitrary initial position and velocities is suggested.
Using this approach we numerically simulated various quasirotational motions
for all the models and obtained that the simple \cite{PRTr} rotational states
of the string model ``triangle" are stable (i.\,e. small disturbances behave
like in the meson model) and the rotational motions of the systems $q$-$q$-$q$
and Y are unstable. In the latter two cases any small asymmetric disturbances
grow with growing time. For the model $q$-$q$-$q$ the middle quark moves away
from the center under the centrifugal force but then it quasiperiodically
returns without merging with an endpoint. The evolution of the three-string
instability includes the complicated motion of the junction and inevitably
results in falling one of the quarks into the junction.

These features of the classical behavior of the string baryon models give some
advantage for the $q$-$qq$ and ``triangle" systems over the $q$-$q$-$q$
and Y configurations. But this does not mean final ``closing"  the latter two
models, in particular, because of the fact, that the majority
of orbitally excited baryon states are resonances so their
classical stability is the problem of minor importance.
So the final choice of the string baryon model is to depend on all
aspects of this problem, including QCD-based grounds \cite{Isg,Corn}
and describing the baryon Regge trajectories \cite{4B,InSh}.

The author are grateful to V.P. Petrov for useful collaboration.
The work is supported by the RBFR  grant  00-02-17359.

\begin{figure}[h]         
\unitlength=1mm
\begin{center}
\begin{picture}(86,8)
\end{picture}\end{center}
\caption{The quasirotational motion of the model $\Delta$.}
\end{figure}

\begin{figure}[h]         
\unitlength=1mm
\begin{center}
\begin{picture}(86,8)
\end{picture}\end{center}
\caption{Behavior of the $q$-$q$-$q$ system (a\,--\,d) in comparison
with the model $\Delta$ (e\,--\,g).}
\end{figure}

\begin{figure}[h]         
\unitlength=1mm
\begin{center}
\begin{picture}(86,8)
\end{picture}\end{center}
\caption{The disturbed motion of the three-string model.}
\end{figure}

\end{document}